\documentclass[
prl,
twocolumn,
showpacs,
superscriptaddress,
nofootinbib
]{revtex4-1}
\usepackage[dvips]{graphicx}
\usepackage{color}
\usepackage{latexsym}
\usepackage{amsmath}
\usepackage{amsthm}
\usepackage{amsfonts}
\usepackage{amssymb}
\usepackage{romannum}
\usepackage[T1]{fontenc}

\usepackage{soul} 			
\usepackage{lineno}
\usepackage[normalem]{ulem}
\usepackage[colorlinks=true,allcolors=blue]{hyperref}
\usepackage[all]{hypcap}

\begin{document}
	
\pagenumbering{arabic}
	
\title{Microwave Spectroscopy Reveals the Quantum Geometric Tensor of Topological Josephson Matter}
\author{R.~L.~Klees}
\affiliation{Fachbereich Physik, Universit{\"a}t Konstanz, D-78457 Konstanz, Germany}
\author{G.~Rastelli}
\affiliation{Fachbereich Physik, Universit{\"a}t Konstanz, D-78457 Konstanz, Germany}
\affiliation{Zukunftskolleg, Universit{\"a}t Konstanz, D-78457 Konstanz, Germany}
\author{J.~C.~Cuevas}
\affiliation{Departamento de F\'{i}sica Te\'{o}rica de la Materia Condensada and Condensed Matter Physics Center (IFIMAC), Universidad Aut\'{o}noma de Madrid, E-28049 Madrid, Spain}
\author{W.~Belzig}
\affiliation{Fachbereich Physik, Universit{\"a}t Konstanz, D-78457 Konstanz, Germany}

\begin{abstract}
	Quantization effects due to topological invariants such as Chern numbers have become very relevant in many systems, yet key quantities such as the quantum geometric tensor providing local information about quantum states remain experimentally difficult to access. 
	Recently, it has been shown that multiterminal Josephson junctions constitute an ideal platform to synthesize topological systems in a controlled manner. 
	We theoretically study properties of Andreev states in topological Josephson matter and demonstrate that the quantum geometric tensor of Andreev states can be extracted by synthetically polarized microwaves. 
	The oscillator strength of the absorption rates provides direct evidence of topological quantum properties of the Andreev states.
\end{abstract}

\date{\today}

\maketitle

\textit{Introduction.}{\textemdash}Presently, there is huge interest in condensed matter physics in topologically nontrivial systems and, in the last two decades, there has been great effort to find novel types of topological quantum matter such as topological insulators \cite{Kane:2005gb,Hasan:2010ku}, topological semimetals \cite{Armitage:2018dg}, or topological superconductors \cite{Sato:2017go}.
The topological phase is often related to isolated singularities in the band structure at which two energy 
bands intersect \cite{Berry:1984ka,Bansil:2016bu}. 
In the case of topological superconductors, Bogoliubov quasiparticles at zero energy, 
called Majorana zero modes, could potentially be used in topologically protected quantum computation  \cite{Sato:2017go}. 
The existence of zero-energy modes in such systems is topologically protected \cite{Fu:2008} which recently has been confirmed in experiments on superconducting three-terminal junctions  \cite{Yang:2019}.
Actually, Andreev bound states (ABS)  in superconducting weak links, also known as Josephson junctions, 
have also been proposed for implementing qubits \cite{Chtchelkatchev:2003jia,Zazunov:2003jm}. 
ABS can be easily tuned if the junctions are embedded in an rf superconducting quantum interference device (SQUID) and  can be experimentally accessed 
and coherently manipulated by microwave \cite{Bretheau:2013by,Janvier:2015fw,vanWoerkom:2017gl,Hays:2018}, tunneling \cite{Pillet:2010}, and supercurrent spectroscopy \cite{Bretheau:2013bt}. 

Recently, multiterminal Josephson junctions (MJJs) made of conventional superconductors have been predicted to exhibit 
nontrivial topology for four \cite{Riwar:2016hr,Yokoyama:2015iu,Yokoyama:2017hv,Eriksson:2017da,Xie:2018kt,Repin:2019ic} and three \cite{vanHeck:2014cm,Padurariu:2015ji,Xie:2017hg,Meyer:2017ge,Gavensky:2019} leads.
In such systems there is no need for exotic topological materials, 
although multiterminal topological nanowires have been discussed as well \cite{Gavensky:2019}.
In MJJs the quantized transconductance across two terminals is a manifestation of the integer-valued 
Chern number \cite{Riwar:2016hr,Eriksson:2017da,Xie:2018kt,Gavensky:2019}. 
Alternatively, Floquet states in periodically driven 
Josephson systems with connectivity simpler than MJJs
can also show nontrivial topology \cite{Gavensky:2018em,Venitucci:2018gb}.
Although it is challenging to fabricate MJJs  \cite{Plissard:2013kk}, a realization of a three-terminal superconducting junction in a double-SQUID configuration and the investigation of its topological properties has already been reported \cite{Strambini:2016gt}.
First experiments towards ballistic MJJs have been performed, too \cite{Draelos:2019gl,Pankratova:2018}.

Since the Chern number follows from integrating the Berry curvature over periodic parameters, 
accessing the more fundamental local properties contained in the quantum geometric tensor (QGT), 
i.e., the Fubini-Study metric tensor and the Berry curvature, provides additional information about the geometry of the state space manifolds \cite{Kolodrubetz:2017jg}.
There have been several proposals how to measure the elements of the QGT, e.g.,
via the noise spectral functions  \cite{Kolodrubetz:2013bf} proposed also for electronic solid state systems \cite{Neupert:2013eu} or via nonadiabatic periodic modulation of the space-defining parameters \cite{Ozawa:2018ky}.
In fact, local topological properties can also be revealed by the quantized spectroscopic response under (nonadiabatic) circular drive \cite{Tran:2017is,Tran:2018jh}, which has already been successfully carried out in Floquet states of ultracold fermionic atoms under time-dependent drive \cite{Flaschner:2016el,Asteria:2019if}.
Similarly, a nontrivial (Floquet) topology was achieved  in superconducting qubits  \cite{Tan:2018dg,Tan:2019}   
generated by custom-built engineered time-dependent drives.

\begin{figure}[t]
	\centering
	\includegraphics[width = \columnwidth]{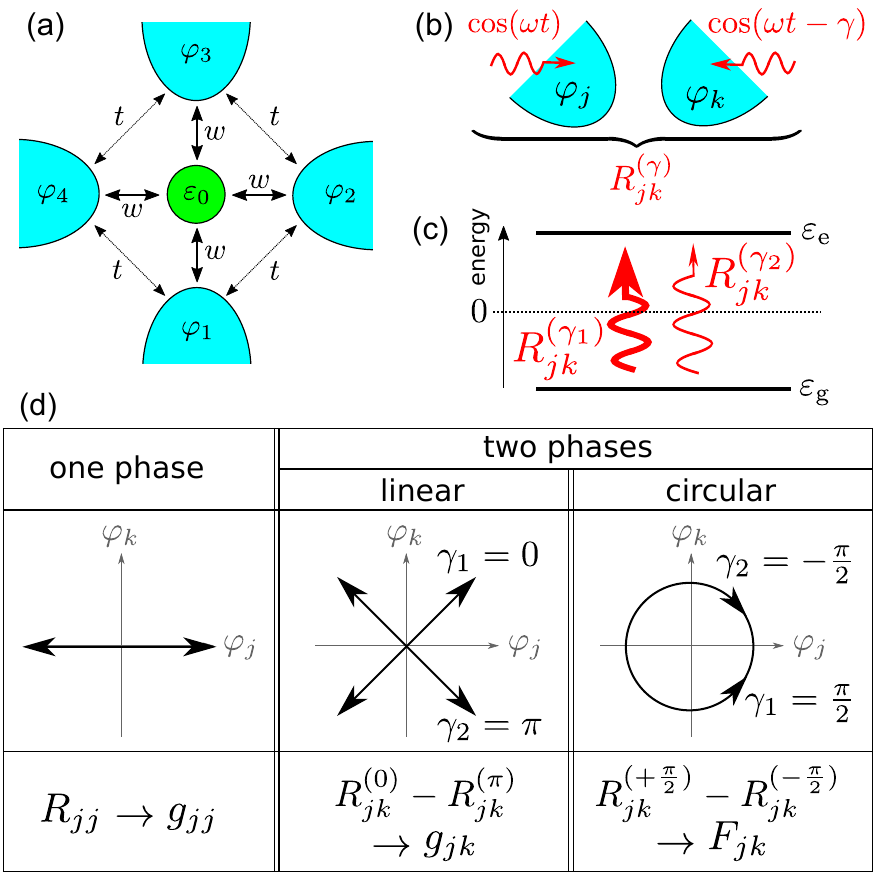}
	\caption{
		Application of polarized microwave spectroscopy in MJJs. 
			(a) Microscopic model of the four terminals. 
			Four superconducting leads, each with a  phase $\varphi_j$ ($j = 1,2,3,4$), are connected to a normal quantum dot with level $\varepsilon_0$ via the couplings $w$.
			The nearest leads are also directly connected to each other by the couplings $t \ll w$.
		(b) A periodic modulation of two phases $\varphi_j$ and $\varphi_k$ ($j \neq k$) at frequency $\omega$ leads to transitions with rates of absorption $R_{jk}^{(\gamma)}$, where $\gamma$ is the relative phase between the two modulations. (c) Two measurements with different relative phases $\gamma_1 \neq \gamma_2$ lead to different transition rates $R_{jk}^{(\gamma_1)} \neq R_{jk}^{(\gamma_2)}$ between the ground state at energy $\varepsilon_\mathrm{g}$ and the excited state at energy $\varepsilon_\mathrm{e}$. (d) Schema of how to extract the elements of the quantum geometric tensor $\chi_{jk} = g_{jk} - i F_{jk}/2$, where $g_{jk}$ is the metric tensor and $F_{jk}$ is the Berry curvature. Driving of one phase $\varphi_j$ allows for the detection of the diagonal elements $g_{jj}$, while linear (circular) driving of two phases $\varphi_j$ and $\varphi_k$ ($j \neq k$) allows for the extraction of the off-diagonal elements $g_{jk}$ ($F_{jk}$).
	}
	\label{fig:fig1}
\end{figure}
In this Letter, we present a straightforward way to experimentally access the full QGT in MJJs.
We show how to extract the elements of the QGT of the ground state manifold of the low-energy ABS hosted in MJJs
by measuring the absorption rates under a weak time-dependent perturbation.
Such linear response measurements have now become standard in ABS spectroscopy \cite{Bretheau:2013by,Janvier:2015fw,vanWoerkom:2017gl}. To this end, we conceive a concrete and feasible experimental way to implement synthetically linear or circular polarized microwave absorption spectroscopy in MJJs.
Figure \ref{fig:fig1} presents the specific example of a four-terminal Josephson junction and 
summarizes the general protocol to extract the full information of the metric tensor and the Berry curvature via microwave absorption spectroscopy.
The latter represents our main result.
Our proposed method can be used in a large variety of topological Josephson matter
and, therefore, provides an unprecedented insight into the nature of quantum states in these systems.
Finally, by integrating the absorption rate differences related to the Berry curvature, our approach allows us to measure the Chern number in MJJs complementary to transconductance measurements.

\textit{Model and effective Hamiltonian.}{\textemdash}For the sake of concreteness, we consider a four-terminal setup sketched in Fig.~\hyperref[fig:fig1]{1(a)} consisting of four superconducting (SC) terminals connected to a normal conducting region that consists of a single level, noninteracting quantum dot providing a spin-degenerate energy level $\varepsilon_0$.
We assume that the SC leads are described by standard BCS-type mean-field Hamiltonians, with a pairing potential $\Delta > 0$ and phase $\varphi_j \in [0,2\pi)$ for $j = 1,\ldots,4$, and that the quantum dot is coupled to the leads with tunneling coupling strength $w > 0$, while the leads are coupled to each other with a tunneling coupling strength $t>0$.

In the large-gap limit $\Delta \to \infty$ and for $t/w \ll 1$, the effective Hamiltonian describing the pair of ABS on the dot reads $H = \Psi^{\dag} H_0 \Psi^{\phantom\dag}$, where $\Psi^{\dag} = (d_\uparrow^\dag, d_\downarrow^{\phantom\dag})$ is the Nambu spinor consisting of an electronic annihilation (creation) $d_\sigma^{(\dag)}$ operator of spin $\sigma$ on the dot and $H_{0} = \boldsymbol{d} \cdot \boldsymbol{\tau}$ describes a pseudospin $\boldsymbol{\tau} = (\tau_1 , \tau_2, \tau_3)^\mathrm{T}$ (Pauli matrices in Nambu space) in an effective magnetic field
\begin{align}
\boldsymbol{d} =  \begin{pmatrix}
	\Gamma \sum_{j=1}^4 \cos\varphi_j \\
	- \Gamma \sum_{j=1}^4 \sin\varphi_j \\
	\varepsilon_{0}   - 2 t_0 \Gamma \sum_{j=1}^4 \cos(\varphi_j - \varphi_{j+1}  )
	\end{pmatrix} 
	\label{eq:vectorD}
\end{align}
controlled by the SC phases $\varphi_j$ \cite{sMaterial}.
Here, $\Gamma = \pi N_0 w^2$ and $t_0 = \pi N_0 t$, where $N_0$ is the normal density of states in the leads at the Fermi level.

\begin{figure}[t]
	\centering
	\includegraphics[width = \columnwidth]{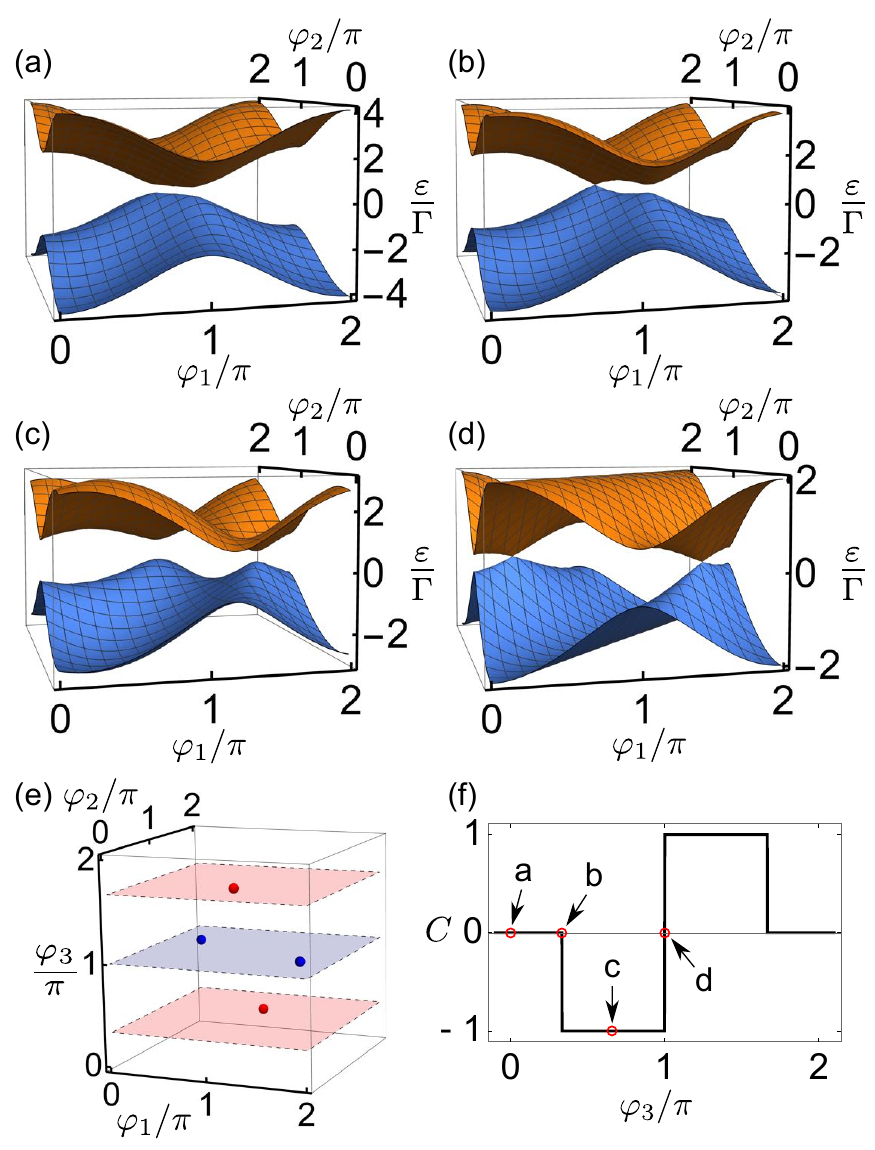}
	\caption{
		Band structure, Weyl nodes and Chern number in the four-terminal junction. 
		(a){\textendash}(d) Energy spectrum $\varepsilon_\mathrm{e/g} = \pm d$ for (a) $\varphi_3 = 0$, (b) $\varphi_3 = \pi/3$, (c) $\varphi_3 = 2\pi/3$, (d) $\varphi_3 = \pi$, respectively. 
		(e) Locations of the four Weyl nodes in the FBZ. Blue (red) Weyl nodes carry a topological charge $c = +1$ ($c = -1$), 
		see Ref.~\cite{sMaterial} for details. 
		The Chern number becomes nontrivial only if the $(\varphi_1,\varphi_2)$ plane of integration lies between two opposing charges. (f) Chern number $C$ as a function of $\varphi_3$. The points a, b, c and d correspond to the values of $\varphi_3$ in panels (a), (b), (c) and (d), respectively. Common parameters for all panels: $t_0 = 0.1$, $\varepsilon_{0}/\Gamma = -0.2$.
	}
	\label{fig:fig2}
\end{figure}

\textit{Andreev states and topology.}{\textemdash}The low-energy Hamiltonian $H_{0}$ defines a two-level system with a ground state (GS) $\left| \mathrm{g} \right\rangle$ and an excited state $\left| \mathrm{e} \right\rangle$, where $H_{0} \left| \mathrm{e/g} \right\rangle = \varepsilon_\mathrm{e/g} \left| \mathrm{e/g} \right\rangle$. 
The pair of ABS has energies given by $\varepsilon_\mathrm{e/g} = \pm d$, where we define $d =  |\boldsymbol{d}|$. 
In the four-terminal case only three SC phases are independent and, therefore, gauge invariance allows us to set one SC phase to zero (from now on we set $\varphi_4 = 0$). 
The remaining three SC phases $\boldsymbol{\varphi} = (\varphi_1,\varphi_2,\varphi_3) \in [ 0 , 2\pi )^3$ define a first Brillouin zone (FBZ) in analogy to the quasimomentum space of a periodic crystal.
The spectrum $\varepsilon_\mathrm{e/g}$ is shown in Figs.~\hyperref[fig:fig2]{2(a){\textendash}2(d)} for several values of $\varphi_3$ which show zero-energy Weyl nodes $\boldsymbol{\varphi}_\mathrm{W}$ separating different gapped phases.
From the constraint $d(\boldsymbol{\varphi}_\mathrm{W})=0$, we find that Weyl nodes only appear if $-8 \leq m \leq 0$ with $m = \varepsilon_0 / t_0 \Gamma$. 
There are four Weyl nodes $\boldsymbol{\varphi}_\mathrm{W}^{(s)} = (\varphi_{\mathrm{W},1}^{(s)},\varphi_{\mathrm{W},2}^{(s)},\varphi_{\mathrm{W},3}^{(s)})$, $s = 1,2,3,4$, located at (modulo $2\pi$ in each direction)
\begin{subequations}
	\label{eq:WeylNodes}
	\begin{align}
		\boldsymbol{\varphi}^{(1)}_\mathrm{W} &= \bigl( -\delta, \pi - \delta, \pi \bigr) ,\\
		\boldsymbol{\varphi}^{(2)}_\mathrm{W} &= \bigl( \delta,  \delta -\pi , \pi \bigr) ,\\
		\boldsymbol{\varphi}^{(3)}_\mathrm{W} &= \bigl( \pi, \pi - \delta, -\delta \bigr) ,\\
		\boldsymbol{\varphi}^{(4)}_\mathrm{W} &= \bigl( \pi, \delta -\pi  , \delta \bigr) ,
	\end{align}
\end{subequations}
where $\delta=\arccos(1 + m/4)$, each of them carrying a topologically positive or negative charge \cite{sMaterial,Wan:2011hi}.
The locations of these zero-energy bound states in the FBZ are shown in Fig.~\hyperref[fig:fig2]{2(e)}.
The existence of well-separated Weyl nodes is robust against small variations of the coupling constants. 
These variations simply move the Weyl points in parameter space away from the locations given in Eq.~\eqref{eq:WeylNodes}  leaving the topological structure of the system intact. Only if the Hamiltonian drastically differs from the presented one, 
Weyl nodes might merge and annihilate. This happens, for instance, if we add a next-nearest-neighbor coupling between leads 1{\textendash}3 and 2{\textendash}4 of the same magnitude as the nearest-neighbor couplings.
We  also remark that the existence of zero-energy solutions $\boldsymbol{\varphi}_\mathrm{W}^{(s)}$ is crucially linked to the presence of a hopping $(t\neq 0)$ 
directly connecting nearest leads. The latter allows for different interfering paths for particles between every two neighboring leads. In the absence of these paths (i.e., $t = 0$), the gap between the ABS cannot be closed for any $\varepsilon_0 \neq 0$ and the system stays topologically trivial.

All SC phases play the role of synthetic $U(1)$ gauge fields for which we define a gauge connection 1-form $\mathcal{A} = \sum_j A_j \, \mathrm{d}\varphi_j$ of the GS $\left| \mathrm{g} \right\rangle$ \cite{footnote1}, where $A_j = i \left\langle \mathrm{g} \middle| \partial_j \mathrm{g} \right\rangle$ is the Berry connection \cite{Berry:1984ka} and $\partial_j \equiv \partial / \partial \varphi_j$. 
The Chern number of the GS manifold is encoded in the gauge-invariant curvature two-form $\mathcal{F} = \mathrm{d}\mathcal{A} = (1/2) \sum_{jk} F_{jk} \, \mathrm{d}\varphi_j \wedge \mathrm{d}\varphi_k$, where $F_{jk} = \partial_j A_k - \partial_k A_j$ is the Berry curvature \cite{Nakahara:2003}.
For our particular two-level Hamiltonian $H_0$, the Berry curvature of the GS in the gapped phase ($d > 0$) can be expressed as $F_{jk}(\boldsymbol{\varphi}) =   \boldsymbol{n} \cdot \bigl[ (\partial_j \boldsymbol{n}) \times  (\partial_k  \boldsymbol{n}) \bigr] /2$ \cite{Hasan:2010ku,sMaterial} via  the normalized effective magnetic field  $\boldsymbol{n} = \boldsymbol{d}/d$.
Defining a Chern number for fixed $\varphi_3$ via
\begin{align}
C(\varphi_3) = \frac{1}{2\pi} \int_0^{2\pi} \!\!\! \mathrm{d}\varphi_1 \int_0^{2\pi} \!\!\! \mathrm{d}\varphi_2 \, F_{12}(\boldsymbol{\varphi}) ,
\label{eq:chernNumber}
\end{align}
we observe topologically nontrivial regions with nonzero Chern number for certain values of $\varphi_3$ [Fig.~\hyperref[fig:fig2]{2(f)}]. 
Depending on the topological charge of a Weyl node, the Chern number changes by $\pm 1$ for each Weyl node that is crossed while moving the $(\varphi_1,\varphi_2)$ plane of integration along the $\varphi_3$ axis. 
Therefore, the finite jumps of $C$ are associated with the values $\varphi_{\mathrm{W},3} = \pi$ and 
$\varphi_{\mathrm{W},3} = \pm \delta$. 
In the shown case for $m = -2$ in Figs.~\hyperref[fig:fig2]{2(e)} and \hyperref[fig:fig2]{2(f)}, the three values of a topological phase transition are $\varphi_{\mathrm{W},3} = \pi / 3, \pi ,  5\pi/3$ in the FBZ. 

\textit{Microwave spectroscopy of quantum geometry.}{\textemdash}The gauge-invariant hermitian QGT of the GS is defined as \cite{Kolodrubetz:2017jg}
\begin{align}
	\chi_{jk} = \left\langle \partial_j \mathrm{g} \middle| \bigl( 1 - \left| \mathrm{g}\right\rangle 
\left\langle\mathrm{g}\right| \bigr) \middle| \partial_k \mathrm{g}\right\rangle\,.
\end{align}
The QGT contains the symmetric (Fubini-Study) metric tensor $g_{jk} = \mathrm{Re}(\chi_{jk})$ measuring the {\textquotedblleft}distance{\textquotedblright} between two adiabatically connected states and the antisymmetric Berry curvature $F_{jk} = -2 \, \mathrm{Im}(\chi_{jk})$ containing information about the geometrical phase acquired during an adiabatic change of parameters. 
Similar to the Berry curvature, also the metric tensor $g_{jk}$ can be conveniently calculated from the normalized effective magnetic field $\boldsymbol{n}$ via $g_{jk} = (\partial_j \boldsymbol{n}) \cdot  (\partial_k \boldsymbol{n}) / 4$  \cite{sMaterial,Shankar2017}.

Let us first show how the diagonal components of the QGT, $\chi_{jj} = g_{jj}$, can be obtained.
For this purpose, we modulate one of the SC phases according to $\varphi_j \to \varphi_j + (2A/\hbar \omega) \cos(\omega t)$, with a frequency $\omega$ and for $(A/\hbar \omega) \ll 1$ [see Fig. \hyperref[fig:fig1]{1(b)}], where $A$ is a coupling parameter, $\hbar$ is Planck's constant, and $t$ is time.
The resulting Hamiltonian to linear order becomes $H = H_0 + 2A (\partial_j H_0) \cos(\omega t) / \hbar \omega$ giving rise to transitions between the two states with 
absorption rates $R_{jj} = r_{jj} \, \delta(2 d  - \hbar\omega)$ by applying Fermi's golden rule [see Fig. \hyperref[fig:fig1]{1(c)}]. The oscillator strength is then given by \cite{Ozawa:2018ky,sMaterial}
\begin{align}
	r_{jj} = \frac{2\pi}{\hbar} A^2 \,    g_{jj} .
	\label{eq:FGRonephase}
\end{align}
The oscillator strength, or the line intensity, can be obtained simply by integration over the
proper frequency range, around $\hbar \omega \approx \varepsilon_\mathrm{e}-\varepsilon_\mathrm{g} = 2 d$. 
Indeed, Eq.~\eqref{eq:FGRonephase} is  valid even in the presence of a finite broadening of the line, as it is expected in microwave experiments.

\begin{figure}[t]
	\centering
	\includegraphics[width = \columnwidth]{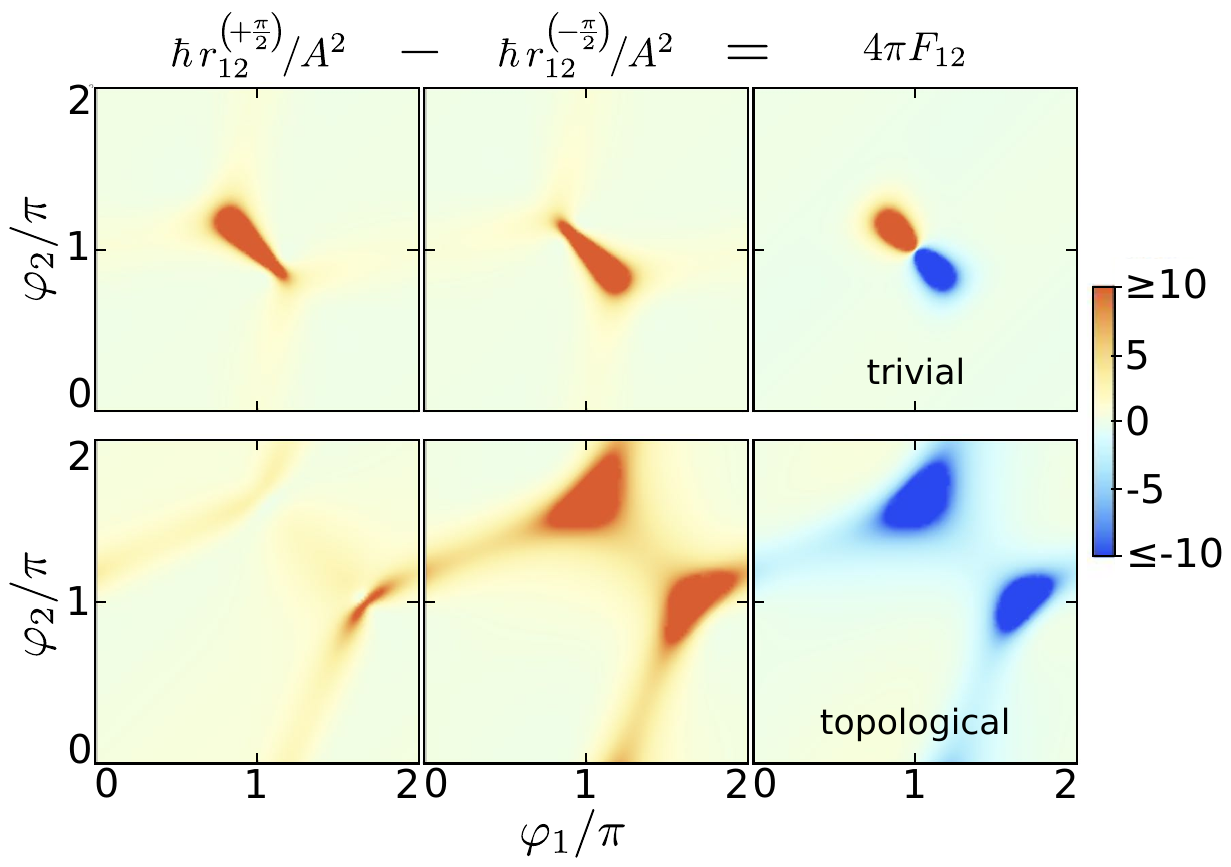}
	\caption{
		Oscillator strengths for right- and left-handed circularly polarized absorption (first and second column). The difference shown in the third column is the Berry curvature [Eq.~\eqref{eq:berryCurvature}]. The upper and lower rows correspond to the trivial ($\varphi_3 = 0$) and the topological ($\varphi_3 = 2\pi/3$) phase, respectively [c.f. Fig.~\hyperref[fig:fig2]{2(f)}]. The parameters are $t_0 = 0.1$, $\varepsilon_{0}/\Gamma = -0.2$.
	}
	\label{fig:fig3}
\end{figure}

Furthermore, the off-diagonal elements are obtained by time-periodic modulation of two phases as shown in Fig.~\hyperref[fig:fig1]{1(b)}.
Depending on the relative phase difference $\gamma$ between both modulations, one obtains the  off-diagonal elements of the QGT $\chi_{jk} = g_{jk} - i F_{jk} / 2$. 
For $j \neq k$, we use the modulations [see Fig. \hyperref[fig:fig1]{1(d)}]
\begin{subequations}
	\begin{align}
	\varphi_j &\to \varphi_j + (2A/\hbar \omega) \cos(\omega t) ,
	\\
	\varphi_k &\to \varphi_k + (2A/\hbar \omega) \cos(\omega t - \gamma) ,
	\end{align}
\end{subequations}
where we again assume $(A/\hbar \omega) \ll 1$. 
As before, we obtain the Hamiltonian to linear order
\begin{align}
	H = H_0 + \frac{2A}{\hbar \omega} \left( \frac{\partial H_0}{\partial \varphi_j} \,  \cos(\omega t)  + 
	\frac{\partial H_0}{\partial \varphi_k} \,  \cos(\omega t - \gamma) \right) ,
\end{align}
from which we obtain the transition absorption rates $R_{jk}^{(\gamma)} = r_{jk}^{(\gamma)} \delta(2d  - \hbar \omega)$ via Fermi's golden rule. The oscillator strength is given by \cite{Ozawa:2018ky,sMaterial}  
\begin{align}
r_{jk}^{(\gamma)}
&= \frac{2\pi  }{\hbar } A^2 \,  (g_{jj} + g_{kk} + 2 g_{jk} \cos\gamma + F_{jk} \sin\gamma) .
\label{eq:rjkgamma}
\end{align}
By performing two subsequent measurements with $\gamma_1 = 0$ and $\gamma_2 = \pi$ (orthogonal linear polarizations), we can extract the off-diagonal part of the metric tensor $g_{jk}$, while two measurements with $\gamma_1 = \pi/2$ and $\gamma_2 = - \pi/2$ (right- and left-handed circular polarization) can be used to measure the Berry curvature $F_{jk}$, i.e.,
\begin{subequations}
	\label{eq:offDiagonalGeometric}
	\begin{align}
		r_{jk}^{(0)} - r_{jk}^{(\pi)}
		&= \frac{8\pi  }{\hbar } A^2 \,  g_{jk} ,
		\label{eq:offDiagonalOfMetric}
		\\
		r_{jk}^{(+ \pi / 2)} - r_{jk}^{(- \pi / 2)}
		&= \frac{4\pi  }{\hbar } A^2 \,  F_{jk}   .
		\label{eq:berryCurvature}
	\end{align}
\end{subequations}
As this gives direct visible evidence about the topological phase of the system, we show the relation between the oscillator strengths for circular drives and the resulting Berry curvatures according to Eq.~\eqref{eq:berryCurvature} for the trivial and the topological phase in Fig.~\ref{fig:fig3} \cite{footnote2}.

Finally, let us recall that once the Berry curvature $F_{jk}$ is extracted, the Chern number $C$ automatically follows from an integration of $F_{jk}$ over the corresponding two SC phases $\varphi_j$ and $\varphi_k$, see Eq.~\eqref{eq:chernNumber}.

\textit{Discussion.}{\textemdash}We have presented a protocol to experimentally measure the QGT of topological Josephson matter via generalized microwave spectroscopy in which different forms of synthetic polarizations are applied. 
The SC phases play the role of quasi-momenta in analogy to topological insulators.
However, the SC phases can be individually fixed and controlled 
by SQUID loops, as achieved in the experiments in Refs.~\cite{Yang:2019} and \cite{Strambini:2016gt}.
The modulation of SC phases can be performed by varying the magnetic fluxes in the SQUID loops with a small ac drive, as reported in the spectroscopy experiments \cite{Bretheau:2013by,Bretheau:2013bt,Janvier:2015fw,vanWoerkom:2017gl}. 

This procedure is not limited to a four-terminal junction, but can be universally applied to any multiterminal Josephson device. 
For instance, another possible realization of topological Josephson matter comprises three SC terminals and the normal region is subjected to a perpendicular magnetic field enclosing a magnetic flux \cite{Meyer:2017ge}. 
This system also supports Weyl nodes and topologically nontrivial states as long as there is a finite direct coupling between the neighboring leads. The low-energy physics on the dot is again described by an effective Hamiltonian of the form $H_{0} = \boldsymbol{d} \cdot \boldsymbol{\tau}$ \cite{sMaterial}. 
The presented microwave protocol can be applied in the same way as before.

We emphasize that our proposed method is an alternative scheme to detect the topological properties beyond the previously suggested transconductance measurements \cite{Riwar:2016hr} with the possible advantage that no electronic contacts are needed. 
A further virtue is that our proposal works at low microwave power such that the linear response regime is applicable. 
We emphasize that MJJs can be intrinsically topological and, hence, do not require the use of designed (eventually strong) time-dependent drives \cite{Gavensky:2018em}.
Because of the universal nature of our proposal, it can be applied to various kinds of topological Josephson matter. 
As long as two SC phases can be addressed independently, the QGT can be determined by engineering the phase difference as we have described. 
It is fair, however, to point out that a realization of the exact setup described in this work certainly requires some
 engineering effort because, in particular, one has to control the coupling between the superconducting leads. Furthermore, 
 the large-gap limit of our model does not consider possible parity jumps due to quasiparticle poisoning \cite{Zgirski:2011dx}.

To conclude, it will be interesting to apply our method to proposed topologically protected candidates for quantum information hardware in superconductors, like, e.g., Majorana states \cite{Kitaev:2007gb} or parafermions \cite{Klinovaja:2014wj}. 
Because of the central quantum dot, our model is an ideal platform to address strong Coulomb interaction and study its effects on both quantum geometry and topology beyond the weak perturbative regime which has already been explored \cite{Marra:2018ch}. However, this goes beyond the scope of the present work.
\newline
\newline
\indent R.L.K. thanks Robert Hussein for useful discussions. This work was supported by the Deutsche Forschungsgemeinschaft through SFB 767 and Grant No. RA 2810/1. J.C.C. acknowledges the support via the Mercator Program of the DFG in the frame of the SFB 767.\\

\noindent \textit{Note added in the proof.}{\textemdash}An experimental measurement of the quantum geometric tensor in qubits formed by NV centers in diamond was recently reported in Ref. \cite{Yu:2019}.

\clearpage

%
%
\renewcommand{\thefigure}{S\arabic{figure}}
\renewcommand{\thetable}{S\arabic{table}}
\renewcommand{\theequation}{S\arabic{equation}}
\renewcommand{\theHtable}{S\arabic{table}}
\renewcommand{\theHfigure}{S\arabic{figure}}
\renewcommand{\theHequation}{S\arabic{equation}}
\setcounter{table}{0}
\setcounter{figure}{0}
\setcounter{equation}{0}
{
\centering
\large Supplemental Material for {\textquotedblleft}Microwave Spectroscopy Reveals the Quantum Geometric Tensor of Topological Josephson Matter{\textquotedblright} \\
\vspace{0.2cm}
\normalsize R. L. Klees, G. Rastelli, J. C. Cuevas, and W. Belzig\\
\href{https://arxiv.org/abs/1810.11277}{arXiv:1810.11277}\\
}
\subsection{Effective low-energy Hamiltonian of multiterminal Josephson junctions}
\begin{figure}
	\includegraphics[width=0.7\columnwidth]{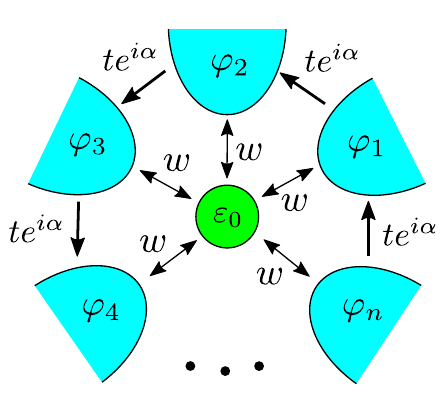}
	\caption{Microscopic model of the $n$-terminal junction. $n$ superconducting leads, each with a superconducting phase $\varphi_j$, are coupled to a normal dot with energy level $\varepsilon_0$ via the couplings $w$. The normal region is subjected a magnetic flux $\Phi = n\alpha \Phi_0$ (flux quantum $\Phi_0 = \hbar / 2 e$) such that the couplings between the superconductors $t$ are modified by a phase factor $e^{i \alpha}$.}
	\label{fig:sfig1}
\end{figure}

We consider $n \in \mathbb{N}$ superconducting (SC) terminals connected to a normal conducting region [see Fig.~\ref{fig:sfig1}] that consists of a single-level, noninteracting quantum dot described by the Hamiltonian $H_\mathrm{D} =  \varepsilon_0 \sum_{\sigma} d_\sigma^\dag d_\sigma^{\phantom\dag}$, where $d_\sigma^\dag$ creates an electron in the dot level with spin $\sigma = \uparrow,\downarrow$ at energy $\varepsilon_0$. 
In the absence of microwave drive, the Hamiltonian of the $n$-terminal junction reads 
\begin{align}
H &=  H_\mathrm{D} + \sum_{j=1}^n \left( H_{\mathrm{S}}^{(j)} +   H_\mathrm{S-D}^{(j)} +  H_\mathrm{S-S}^{(j,j+1)} \right) ,
\end{align}
with
\begin{subequations}
	\begin{align}
	H_{\mathrm{S}}^{(j)} &= \sum\nolimits_{\boldsymbol{k}\sigma} \xi_{\boldsymbol{k}} \,  c_{j \boldsymbol{k} \sigma}^{\dag} c_{j \boldsymbol{k} \sigma}^{\phantom\dag}  
	\nonumber \\
	&\quad + \Delta  \sum\nolimits_{\boldsymbol{k}}  \Bigl[ \, e^{i \varphi_j} c_{j \boldsymbol{k}\uparrow}^\dag c_{j (-\boldsymbol{k})\downarrow}^{\dag} + \mathrm{H.c.} \Bigr] ,
	\\
	H_\mathrm{S-D}^{(j)} &= \sum\nolimits_{\boldsymbol{k}\sigma} w \Bigl[ c_{j \boldsymbol{k}\sigma}^\dag d_{ \sigma}^{\phantom\dag} + d_{ \sigma}^{\dag} 
	c_{j \boldsymbol{k} \sigma}^{\phantom\dag} \Bigl] ,
	\\
	H_\mathrm{S-S}^{(j,j+1)} &= 
	\sum\nolimits_{\boldsymbol{k} \sigma} t \Bigl[ e^{-i\alpha} c_{j \boldsymbol{k} \sigma}^\dag c_{(j+1) \boldsymbol{k} \sigma}^{\phantom\dag} + \mathrm{H.c.} \Bigl],
	\end{align}
\end{subequations}
where $H_{\mathrm{S}}^{(j)}$ is the Hamiltonian of the $j$-th SC lead, $H_\mathrm{S-D}^{(j)}$ is the coupling between the $j$-th SC lead and the dot with coupling strength $w$ and $H_\mathrm{S-S}^{(j,j+1)}$ describes the coupling between neighboring SC leads (periodic: $j = n 
\Rightarrow j+1 = 1$) which is assumed to be weak, i.e., $t \ll w$.
We furthermore include the effect of a magnetic flux $\Phi = n\alpha\Phi_0$ in the normal region with $\Phi_0 = \hbar / 2e$ being the flux quantum leading to a hopping phase $\alpha$ between two neighboring leads.
Moreover, $c_{j \boldsymbol{k} \sigma}^\dag$ creates an electronic state in lead 
$j$ with quasi-momentum $\boldsymbol{k}$ and spin $\sigma = \uparrow,\downarrow$ at energy $\xi_{\boldsymbol{k}} = \hbar^2\boldsymbol{k}^2 / 2 m_\mathrm{e} - \mu$, where $m_\mathrm{e}$ is the mass of the electrons and $\mathrm{H.c.}$ denotes the Hermitian conjugate.
We assume that all leads have the same absolute value of the SC gap $\Delta$, the same chemical potential $\mu$, but they differ in the SC phase denoted by $\varphi_j$. 
By introducing spinors in Nambu space $\Psi_{j \boldsymbol{k}}^{\dag} = (c_{j \boldsymbol{k} \uparrow}^\dag, c_{j (-\boldsymbol{k}) \downarrow})$ for the leads ($j = 1,\ldots,n$) and $\Psi_\mathrm{D}^{\dag} = (d_\uparrow^\dag, d_\downarrow^{\phantom\dag})$, we rewrite the unperturbed Hamiltonians as $H_{\mathrm{S}}^{(j)} = \sum_{\boldsymbol{k}} \Psi_{j \boldsymbol{k}}^{\dag} \hat{H}_{\mathrm{S},j\boldsymbol{k}} \Psi_{j \boldsymbol{k}}^{\phantom\dag}$ and $H_{\mathrm{D}}= \Psi_\mathrm{D}^{\dag} \hat{H}_\mathrm{D} \Psi_\mathrm{D}^{\phantom\dag}$, respectively, with $\hat{H}_{\mathrm{S},j\boldsymbol{k}} = \xi_{\boldsymbol{k}} \tau_3 + \Delta e^{i \varphi_j \tau_3} \tau_1$ and $\hat{H}_\mathrm{D} = \varepsilon_0 \tau_3$ and the Pauli matrices  $\tau_1,\tau_2$ and $\tau_3$ in Nambu space. 
Note that due to isotropic symmetry $\xi_{\boldsymbol{k}} = \xi_{-\boldsymbol{k}}$ holds, which we have already used.

From $\hat{g}_{j\boldsymbol{k}} = (\varepsilon - \hat{H}_{\mathrm{S},j\boldsymbol{k}})^{-1}$, we obtain the bare Green's function (GF) of the $j$-th lead
\begin{align}
\hat{g}_{j\boldsymbol{k}} = \frac{\varepsilon - \xi_{\boldsymbol{k}} \tau_3 + \Delta e^{i \varphi_j \tau_3} \tau_1}{\varepsilon^2 - \xi_{\boldsymbol{k}}^2 - \Delta^2}.
\end{align}
Since the tunneling matrix elements $t$ and $w$ do not depend on quasi-momentum $\boldsymbol{k}$, we can already sum over all quasi-momenta to obtain the new GF of the lead $\hat{g}_{j} = \sum\nolimits_{\boldsymbol{k}} \hat{g}_{j\boldsymbol{k}}$. 
Turning the summation over $\boldsymbol{k}$ into an integration over energies in the wide-band limit via $\sum_{\boldsymbol{k}} \to N_0 \int \mathrm{d}\xi_{\boldsymbol{k}}$, where $N_0$ is the normal state density of states at the Fermi energy, we obtain
\begin{align}
\hat{g}_{j} &= - \pi N_0 \frac{\varepsilon  + \Delta e^{i \varphi_j \tau_3} \tau_1}{ \sqrt{\Delta^2 - \varepsilon^2 } } \, \stackrel{\Delta \to \infty}{\longrightarrow} \, - \pi N_0  e^{i \varphi_j \tau_3} \tau_1 
\end{align}
defined in the local basis $\Psi_{j}^{\dag} = (c_{j  \uparrow}^\dag, c_{j \downarrow})$, where we assume $\Delta$ to be larger than all relevant energies in the system. 
The GF of the bare dot is given by $\hat{g}_\mathrm{D} = 1 / (\varepsilon - \varepsilon_0 \tau_3)$. 
From the Hamiltonian describing the hopping between the $j$-th lead and the dot we obtain $\hat{V}_{j\mathrm{D}} = w \tau_3$, while we obtain for the coupling between the $j$-th and $(j+1)$-th lead $\hat{V}_{j(j+1)} = t \tau_3 e^{i \alpha \tau_3}$ (periodic: $j = n$ $\Rightarrow$ $j+1 = 1$).
By means of Dyson's equation $G = g + g V G$, where $G$ is the dressed GF of the total system, we find the system of coupled equations
\begin{subequations}
	\begin{align}
	\hat{G}_\mathrm{D} &= \hat{g}_\mathrm{D} + w \, \hat{g}_\mathrm{D} \tau_3 
	\sum_{j=1}^{n}  \hat{G}_{j \mathrm{D}} ,
	\\
	\hat{G}_{j \mathrm{D}} &= \hat{g}_\mathrm{D} 
	\bigl( \hat{V}_{j\mathrm{D}}  \hat{G}_\mathrm{D} 
	+ \hat{V}_{j,j-1}  \hat{G}_\mathrm{j-1,D} 
	+ \hat{V}_{j,j+1}  \hat{G}_\mathrm{j+1,D} 
	\bigr) ,
	\\
	\hat{G}_{j\pm 1,\mathrm{D}} &= \hat{g}_{j \pm 1,j \pm 1} \sum_k \hat{V}_{j \pm 1,k}  \hat{G}_{k\mathrm{D}} ,
	\end{align}
\end{subequations}
which we solve for the dressed GF of the dot to linear order in $t$ and obtain
\begin{align}
\hat{G}_\mathrm{D}^{-1} = \hat{g}_\mathrm{D}^{-1} -  \hat{\Sigma}_\mathrm{D},
\end{align}
with the self-energy
\begin{align}
\hat{\Sigma}_\mathrm{D} &= \sum_{j=1}^n \Bigl( 
\hat{V}_{\mathrm{D} j} \hat{g}_{jj}  \hat{V}_{j \mathrm{D} } 
+
\hat{V}_{\mathrm{D} j} \hat{g}_{jj}  \hat{V}_{j,j-1 } \hat{g}_{j-1,j-1 } \hat{V}_{j-1,\mathrm{D} }
\nonumber \\
&\quad +
\hat{V}_{\mathrm{D} j} \hat{g}_{jj}  \hat{V}_{j,j+1 } \hat{g}_{j+1,j+1 } \hat{V}_{j+1,\mathrm{D} }
\Bigr) + \mathcal{O}(t^2) .
\end{align}
From the self-energy, we obtain the effective low-energy Hamiltonian $H_0$ of the dot as
\begin{align}
H_0 = \varepsilon - \hat{G}_\mathrm{D}^{-1} = \varepsilon - \hat{g}_\mathrm{D}^{-1} + \hat{\Sigma}_\mathrm{D} = \varepsilon_0 \tau_3 + \hat{\Sigma}_\mathrm{D} ,
\end{align}
which can be written as $H_0 = \boldsymbol{d} \cdot \boldsymbol{\tau}$, with the pseudo-spin $\boldsymbol{\tau}$ (set of Pauli matrices in Nambu space) and the effective magnetic field 
\begin{align}
\boldsymbol{d} =  \begin{pmatrix}
\Gamma \sum_{j=1}^n \cos\varphi_j \\
- \Gamma \sum_{j=1}^n \sin\varphi_j \\
\varepsilon_{0}   - 2 t_0 \Gamma \sum_{j=1}^n \cos(\varphi_j - \varphi_{j+1} -\alpha  )
\end{pmatrix} ,
\label{eq:magField}
\end{align}
where we defined $\Gamma = \pi N_0 w^2$ and $t_0 = \pi N_0 t$. The case $n=4$ and $\alpha = 0$ is discussed in the main text, see Eq.~\eqref{eq:vectorD}.
This two-level Hamiltonian has a ground state $\left| \mathrm{g} \right\rangle $ and an excited state $\left| \mathrm{e} \right\rangle$ which satisfy $H_0 \left| \mathrm{e/g} \right\rangle = \varepsilon_{\mathrm{e/g}} \left| \mathrm{e/g} \right\rangle$, with the eigenenergies $\varepsilon_{\mathrm{e/g}} = \pm d$ and the absolute value of the effective field $d = \sqrt{d_1^2 + d_2^2 + d_3^2}$. 
The eigenstates are given by
\begin{align}
\left| \mathrm{e/g} \right\rangle 
&= \frac{1}{\sqrt{2d (d\mp d_3 )}} 
\begin{pmatrix}
d_1 - i d_2  \\
\pm d - d_3
\end{pmatrix}
\nonumber \\
&= \frac{1}{\sqrt{2 \mp 2 \cos\theta}} 
\begin{pmatrix}
e^{-i \varphi } \sin\theta \\
\pm 1 - \cos\theta
\end{pmatrix} ,
\label{eq:GroundExcitedState}
\end{align}
where the spherical angles $\theta \in [0,\pi)$ and $\varphi \in [0,2\pi)$ are defined by the set of equations
\begin{align}
\cos\varphi &= \frac{d_1}{\sqrt{d_1^2 + d_2^2}} , 
\quad  \cos\theta = \frac{d_3}{d } , 
\nonumber \\
\sin\varphi &= \frac{d_2}{\sqrt{d_1^2+ d_2^2}}, 
\quad \sin\theta = \frac{\sqrt{d_1^2 + d_2^2}}{d } ,
\label{eq:relationsAnglesToParameters}
\end{align}
which parametrize the Bloch sphere. 
This explicitly shows that the eigenstates are independent of the absolute value $d$ of the effective field.
\subsection{Topological charge of Weyl points}
Here, we describe how to obtain the sign of the topological charge of the four Weyl points. 
We focus on the case $n=4$ and $\alpha = 0$ as discussed in the main text. 
In the vincinity of one of the points of degeneracy $\boldsymbol{\varphi}_\mathrm{W}^{(s)} = (\varphi_{\mathrm{W},1}^{(s)},\varphi_{\mathrm{W},2}^{(s)},\varphi_{\mathrm{W},3}^{(s)})$ for $s = 1,2,3,4$, we linearize the spectrum via the replacement $\varphi_{k} = \varphi_{\mathrm{W},k}^{(s)} + \delta\varphi_{k}$ for $k = 1,2,3$, where $\delta\varphi_{k}$ is a small deviation from the Weyl point in the direction of $\varphi_{k}$. 
Since for all Weyl points $d(\boldsymbol{\varphi}_\mathrm{W}^{(s)}) = 0$ holds, we obtain $\boldsymbol{d} = \boldsymbol{M}^{(s)} \boldsymbol{\delta\varphi}$ with $\boldsymbol{\delta\varphi} = (\delta\varphi_1 , \delta\varphi_2 , \delta \varphi_3)^\mathrm{T}$ and the matrix elements 
\begin{subequations}
	\begin{align}
	M_{1k}^{(s)} &= -  \Gamma \sin  \varphi_{\mathrm{W},k}^{(s)} ,  \\
	M_{2k}^{(s)} &= - \Gamma \cos  \varphi_{\mathrm{W},k}^{(s)} , \\
	M_{31}^{(s)} &= 2\Gamma t_0 [ \sin(\varphi_{\mathrm{W},1}^{(s)} - \varphi_{\mathrm{W},2}^{(s)}) +  \sin(\varphi_{\mathrm{W},1}^{(s)}) ] , \\
	M_{32}^{(s)} &=2 \Gamma t_0 [  \sin(\varphi_{\mathrm{W},2}^{(s)} - \varphi_{\mathrm{W},3}^{(s)}) - \sin(\varphi_{\mathrm{W},1}^{(s)} - \varphi_{\mathrm{W},2}^{(s)}) ] , \\
	M_{33}^{(s)} &= 2 \Gamma t_0 [ \sin(\varphi_{\mathrm{W},3}^{(s)}) -  \sin(\varphi_{\mathrm{W},2}^{(s)} - \varphi_{\mathrm{W},3}^{(s)})   ] ,
	\end{align}
\end{subequations}
of the transformation matrix $\boldsymbol{M}^{(s)} = (M_{jk}^{(s)})$ which define the {\textquotedblleft}velocities{\textquotedblright} $\boldsymbol{v}_k^{(s)} = (M_{k1}^{(s)},M_{k2}^{(s)},M_{k3}^{(s)})^\mathrm{T}$. 
The Weyl-Hamiltonian can then be written as
\begin{align}
H_\mathrm{W}^{(s)} &=  \sum\nolimits_{kl}  \tau_l M_{lk}^{(s)} \, \delta\varphi_k   = \sum\nolimits_{k}   (\boldsymbol{v}_k^{(s)} \cdot \boldsymbol{\delta\varphi}  ) \tau_k .
\end{align}
The topological charge $c_s$ (or chirality) of a Weyl point is defined via $c_s = \mathrm{sgn}[\boldsymbol{v}_1^{(s)} \cdot (\boldsymbol{v}_2^{(s)} \times \boldsymbol{v}_3^{(s)})] = \mathrm{sgn}[\mathrm{det}(\boldsymbol{M}^{(s)})]$, where the triple product of the velocities is
\begin{align}
&\mathrm{det}(\boldsymbol{M}^{(s)}) =  2 t_0 \Gamma^3 \sin(\varphi_{\mathrm{W},1}^{(s)} - \varphi_{\mathrm{W},3}^{(s)})  
\nonumber \\
& \quad \bigl[ \sin(\varphi_{\mathrm{W},1}^{(s)} - \varphi_{\mathrm{W},2}^{(s)}) + \sin(\varphi_{\mathrm{W},2}^{(s)})  - \sin(\varphi_{\mathrm{W},2}^{(s)} - \varphi_{\mathrm{W},3}^{(s)}) \bigr] .
\label{eq:topCharge}
\end{align}
From Eq.~\eqref{eq:WeylNodes} in the main text, we see that we have four Weyl points in the 3D parameter space, where $\boldsymbol{\varphi}_\mathrm{W}^{(1,2)}$ are located in the $\varphi_3 = \pi$ plane and $\boldsymbol{\varphi}_\mathrm{W}^{(3,4)}$
are located in the $\varphi_1 = \pi$ plane and recall that $-8 < m < 0$ and, therefore, $\mathrm{sgn}(m) = -1$. 
For the four points we find $c_{1,2} =  \mathrm{sgn}(t_0) $ and $c_{3,4} = - \mathrm{sgn}(t_0)$ since $\Gamma > 0$.
This also follows from time-reversal symmetry which links $\boldsymbol{\varphi}_\mathrm{W}^{(1)} = - \boldsymbol{\varphi}_{\mathrm{W}}^{(2)} (+ 2 \pi \boldsymbol{z})$, $\boldsymbol{z} \in \mathbb{Z}^3$, and $\boldsymbol{\varphi}_{\mathrm{W}}^{(3)} = - \boldsymbol{\varphi}_{\mathrm{W}}^{(4)} (+ 2 \pi \boldsymbol{z})$, $\boldsymbol{z} \in \mathbb{Z}^3$. 
Weyl points which are linked by time-reversal symmetry must have the same charge.
\subsection{Quantum geometric tensor from an effective magnetic field}
\begin{figure*}
	\includegraphics[width=\textwidth]{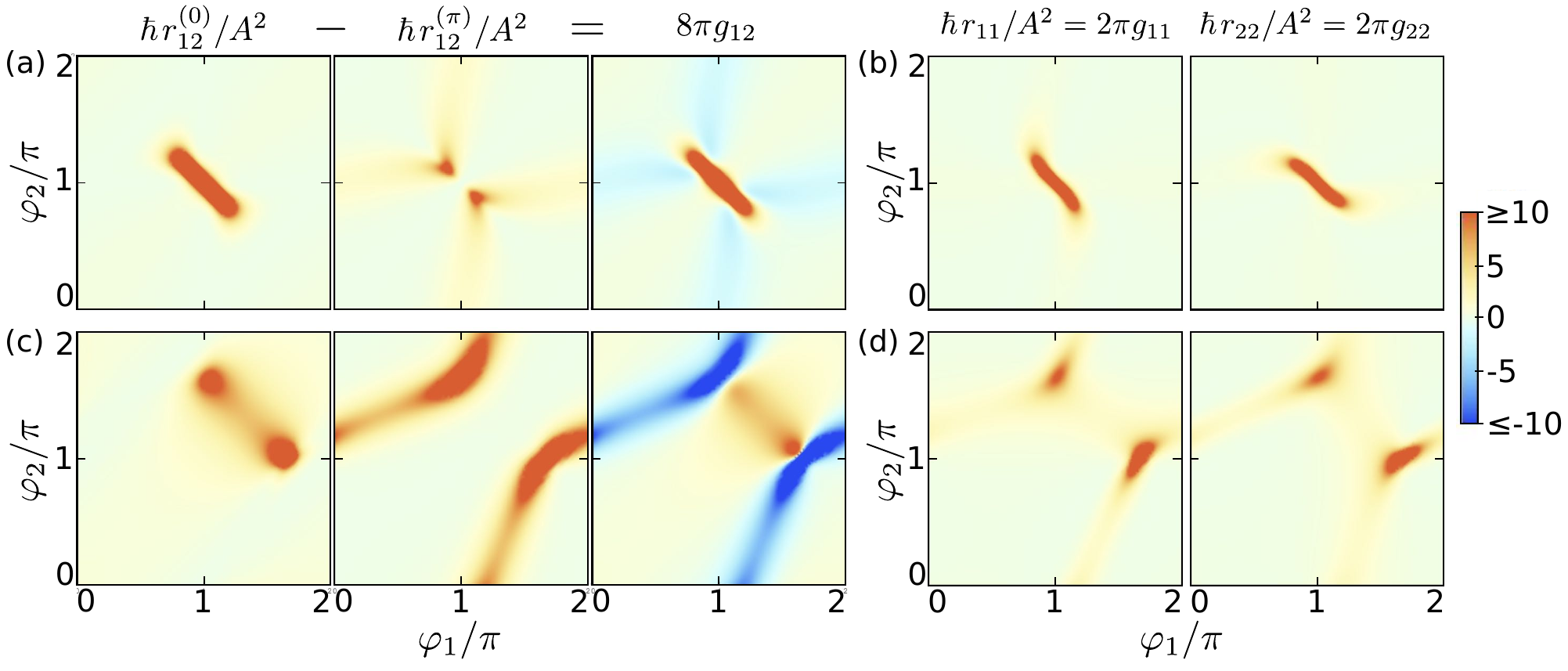}
	\caption{Oscillator strengths for linearly polarized absorption and resulting elements of the metric tensor [Eqs.~\eqref{eq:FGRonephase} and \eqref{eq:offDiagonalOfMetric} in the main text]. (a) Off-diagonal and (b) diagonal elements in the trivial phase ($\varphi_3 = 0$). (c) Off-diagonal and (d) diagonal elements in the topological phase ($\varphi_3 = 2\pi/3$). The phases are defined by the Chern number, see Fig.~\hyperref[fig:fig2]{2(f)} in the main text. Parameters: $t_0 = 0.1, \varepsilon_0 / \Gamma = -0.2$. }
	\label{fig:sfig3}
\end{figure*}
Here, we provide explicit formulas of how to compute the Berry curvature $F_{jk}$ and the quantum metric tensor $g_{jk}$ of the ground state $\left| \mathrm{g} \right\rangle $ directly from the effective Hamiltonian $H_0 = \boldsymbol{d} \cdot  \boldsymbol{\tau}$. 
In general, the effective field is controlled by a set of $n$ parameters $ \boldsymbol{\lambda}  = (\lambda_1, \ldots , \lambda_n)$, i.e., $ \boldsymbol{d}  =  \boldsymbol{d} (\boldsymbol{\lambda})$. 
Using the explicit expression of $\left| \mathrm{g} \right\rangle $, see Eq.~\eqref{eq:GroundExcitedState}, in the definition of the quantum geometric tensor of the ground state $\chi_{jk} = \left\langle  \partial_j \mathrm{g} \middle| (1 - \left| \mathrm{g} \right\rangle \left\langle \mathrm{g} \right| ) \middle| \partial_k \mathrm{g} \right\rangle$, we find
\begin{align}
\chi_{jk} &= \frac{1}{4} \bigl( \partial_j \theta(\boldsymbol{\lambda})  + i \sin\theta(\boldsymbol{\lambda}) \, \partial_j \varphi(\boldsymbol{\lambda})    \bigr)
\nonumber \\
&\quad \times 
\bigl( \partial_k \theta(\boldsymbol{\lambda}) - i \sin\theta(\boldsymbol{\lambda}) \, \partial_k \varphi(\boldsymbol{\lambda}) \bigr),
\end{align}
where $\partial_j = \partial/\partial \lambda_j$. 
The quantum metric (i.e., the Fubini-Study metric) is given by the real part, $g_{jk} = \mathrm{Re}(\chi_{jk})$, and the Berry curvature is given by the imaginary part, $F_{jk} = - 2  \mathrm{Im}(\chi_{jk})$. 
We obtain
\begin{subequations}
	\begin{align}
	F_{jk} &= \frac{1}{2}  \sin\theta(\boldsymbol{\lambda}) \left(
	\frac{\partial \theta(\boldsymbol{\lambda})}{\partial\lambda_j}  \frac{\partial \varphi(\boldsymbol{\lambda})}{\partial\lambda_k}
	-  \frac{\partial \theta(\boldsymbol{\lambda})}{\partial\lambda_k} \frac{\partial \varphi(\boldsymbol{\lambda})}{\partial\lambda_j} 
	\right) ,
	\label{eq:berryC}
	\\
	g_{jk} &= \frac{1}{4} 
	\left(  
	\frac{\partial \theta(\boldsymbol{\lambda})}{\partial\lambda_j} \frac{\partial \theta(\boldsymbol{\lambda})}{\partial\lambda_k}
	+  \sin^2\theta(\boldsymbol{\lambda}) \,  \frac{\partial \varphi(\boldsymbol{\lambda})}{\partial\lambda_j} \frac{\partial \varphi(\boldsymbol{\lambda})}{\partial\lambda_k}
	\right) . 
	\label{eq:metric}
	\end{align}
\end{subequations}
We find by differentiating Eq.~\eqref{eq:relationsAnglesToParameters} with respect to $\lambda_j$:
\begin{subequations}
	\label{eq:derivatives}
	\begin{align}
	\frac{\partial \varphi(\boldsymbol{\lambda})}{\partial\lambda_j}   &
	=  \frac{1}{d_1^2 + d_2^2 }
	\sum_{\alpha,\beta=1}^2 \varepsilon_{\alpha\beta} d_\alpha \partial_j d_\beta ,
	,
	\\
	\frac{\partial \theta(\boldsymbol{\lambda})}{\partial\lambda_j}   
	&= \frac{1}{ d^2 \sqrt{d_1^2 + d_2^2 }} \sum_{\alpha=1}^2 d_\alpha   
	\left(  d_3 \partial_j d_\alpha  -  d_\alpha   \partial_j d_3  \right) ,
	\end{align}
\end{subequations}
where $\varepsilon_{\alpha\beta} = - \varepsilon_{\beta\alpha}$ is the total antisymmetric Levi-Civita tensor in two dimensions, with $\varepsilon_{12} = 1$. 
Using Eq.~\eqref{eq:derivatives} in Eq.~\eqref{eq:berryC}, we find that, after some algebra, the Berry curvature can be written as
\begin{align}
F_{jk} 
&=  \frac{ \boldsymbol{d} \cdot [    (\partial_j \boldsymbol{d}) \times (\partial_k \boldsymbol{d})  ]}{2d^3} 
\nonumber \\
&=  \frac{1 }{2} \Bigl(  \boldsymbol{n} \cdot [(\partial_j  \boldsymbol{n}) \times  (\partial_k  \boldsymbol{n})] 
\Bigr) ,
\end{align}
which allows for the computation of the Berry curvature directly from the effective magnetic field $\boldsymbol{d}$. 
For the second equation we introduced the normaized effective field $\boldsymbol{n} = \boldsymbol{d} / d$ which shows explicitly that the Berry curvature is independent of the absolute value $d$. 
A similar formula can be obtained for the quantum metric tensor. 
Using Eq.~\eqref{eq:derivatives} in Eq.~\eqref{eq:metric}, we find after some algebra
\begin{align}
g_{jk} &= \frac{1}{4d^4}  \biggl(     
d^2 \Bigl[(\partial_j \boldsymbol{d}) \cdot (\partial_k \boldsymbol{d}) \Bigr] - \Bigl[ \boldsymbol{d} \cdot (\partial_j \boldsymbol{d})\Bigr] \ \Bigl[\boldsymbol{d} \cdot (\partial_k \boldsymbol{d})\Bigr] 
\biggr) 
\nonumber \\
&= \frac{1}{4} (\partial_j \boldsymbol{n}) \cdot  (\partial_k \boldsymbol{n}) .
\end{align}
\subsection{Quantum geometric tensor and transition rates}
\begin{figure*}
	\includegraphics[width=\textwidth]{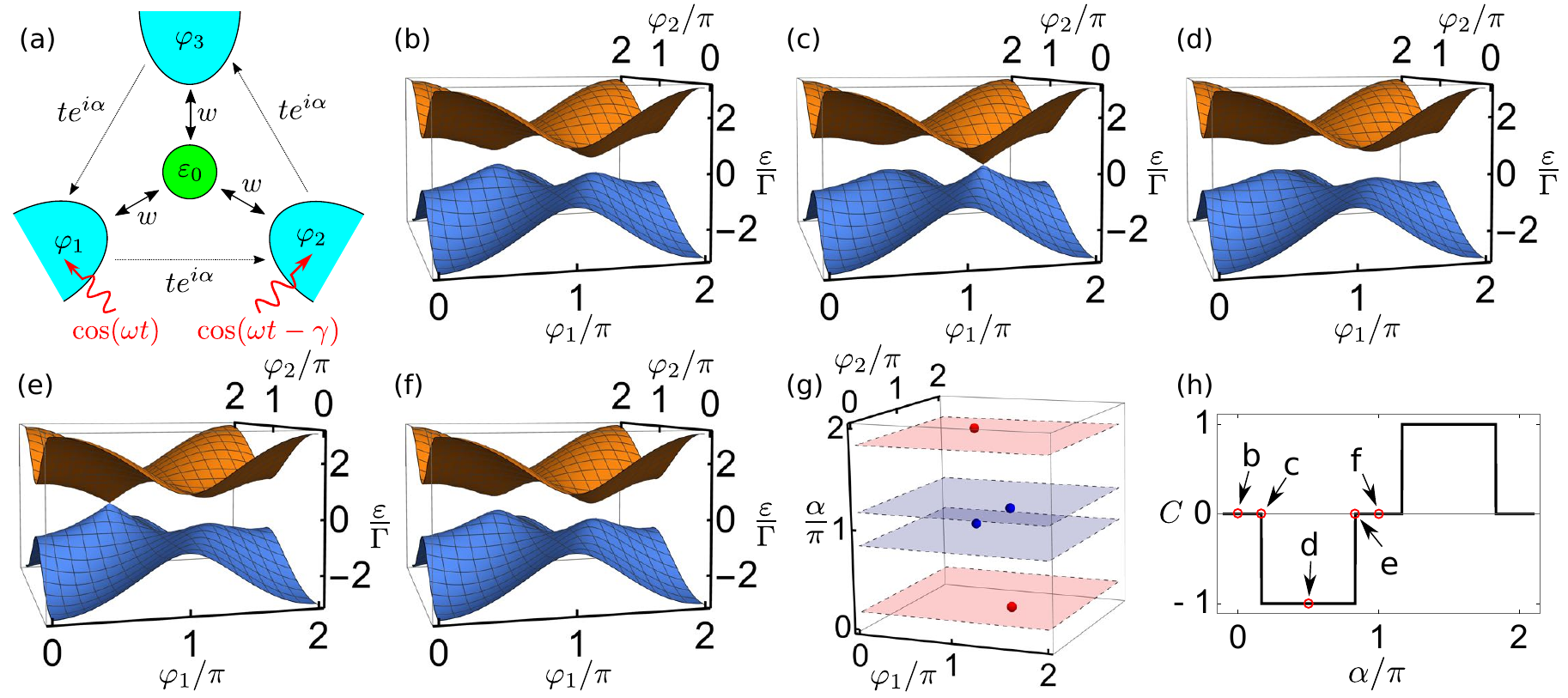}
	\caption{Model, band structure, Weyl nodes and Chern number in a three-terminal Josephson junction. (a) Microscopic model shown in Fig.~\ref{fig:sfig1} for $n=3$ terminals. Red wiggly arrows indicate that up to two superconducting phases can be modulated at frequency $\omega$ and relative phase difference $\gamma$ to obtain information about the quantum geometric tensor as explained in the main text. (b){\textendash}(f) Energy spectrum $\varepsilon_\mathrm{e/g} = \pm d$ for (b) $\alpha = 0$, (c) $\alpha = \pi/6$, (d) $\alpha = \pi/2$, (e) $\alpha = 5\pi/6$, (f) $\alpha = \pi$,  respectively. (g) Locations of the four Weyl nodes in the FBZ. Blue (red) Weyl nodes carry a topological charge $c = +1$ ($c = -1$). The Chern number becomes nontrivial only if the $(\varphi_1,\varphi_2)$ plane of integration lies between two opposing charges. (h) Chern number $C$ as a function of $\alpha$. The points b, c, d, e, and f correspond to the values of $\alpha$ in panels (b), (c), (d), (e) and (f), respectively. Common parameters for all panels: $t_0 = 0.1$, $\varepsilon_0 / \Gamma = 0$.}
	\label{fig:sfig2}
\end{figure*}
The effective low-energy Hamiltonian of the dot in all the realizations of topological Josephson matter discussed in this work adopts the form $H_{0} = \boldsymbol{d} \cdot \boldsymbol{\tau}$. 
This Hamiltonian describes an effective two-level system with $H_0 \left| \mathrm{e/g} \right\rangle = \varepsilon_\mathrm{e/g} \left| \mathrm{e/g} \right\rangle$, where $\varepsilon_\mathrm{g} = -d$ and $\varepsilon_\mathrm{e} = +d$ are the Andreev bound state energies of the ground state (GS) $\left| \mathrm{g} \right\rangle$ and the excited state $\left| \mathrm{e} \right\rangle$, respectively. 
Here, $d = \sqrt{d_1^2 + d_2^2 + d_3^2}$ denotes the absolute value of $\boldsymbol{d}$.

From $\left| \mathrm{g} \right\rangle$, we calculate the  Berry connection $A_j = i \left\langle \mathrm{g} \middle| \partial_j \mathrm{g} \right\rangle$ and the Berry curvature 
\begin{align}
F_{jk} &= \partial_j A_k - \partial_k  A_j 
= i \left[ \left\langle \partial_j \mathrm{g} \middle| \partial_k \mathrm{g} \right\rangle -  \left\langle \partial_k \mathrm{g} \middle| \partial_j \mathrm{g} \right\rangle \right]
\end{align}
of the GS where now $\partial_j = \partial / \partial \varphi_j$.
Furthermore, we need the relation
\begin{align}
0 =  \left\langle \mathrm{e} \middle| (\partial_j H_0) \middle| \mathrm{g} \right\rangle + 2 d \left\langle \mathrm{e} \middle| \partial_j  \mathrm{g} \right\rangle ,
\label{eq:sm22}
\end{align}
which follows from $\partial_j\left\langle \mathrm{e} \middle|  H_0 \middle| \mathrm{g} \right\rangle=0$. 
First, we proof Eq.~\eqref{eq:FGRonephase} for the oscillator strength in the main text. According to Fermi's golden rule, the transition rate due to a perturbation $2A (\partial_j H_0) \cos(\omega t) / \hbar \omega$ is given by
\begin{align}
R_{jj} = \frac{2\pi}{\hbar} \frac{A^2}{(\hbar\omega)^2} \left| \left\langle \mathrm{e} \right| (\partial_j H_0)  \left| \mathrm{g} \right\rangle \right|^2 \delta(2d - \hbar \omega) .
\end{align}
We are left to evaluate the matrix element which reads
\begin{align}
\left| \left\langle \mathrm{e} \right| (\partial_j H_0)  \left| \mathrm{g} \right\rangle \right|^2
&\stackrel{\eqref{eq:sm22}}{=} 
4 d^2 
\left\langle \partial_j  \mathrm{g} \middle| \mathrm{e}  \right\rangle
\left\langle \mathrm{e} \middle| \partial_j  \mathrm{g} \right\rangle 
= 4 d^2 
g_{jj}
\end{align}
since $\left|\mathrm{g} \right\rangle \left\langle \mathrm{g} \right| + \left|\mathrm{e} \right\rangle \left\langle \mathrm{e} \right| = 1$. 

Now, we proof Eqs.~\eqref{eq:rjkgamma} and \eqref{eq:offDiagonalGeometric} in the main text relating the off-diagonal elements of the quantum geometric tensor to the oscillator strengths. 
According to Fermi's golden rule, the transition rate due to a perturbation $2A (\partial_j H_0) \cos(\omega t) / \hbar \omega + 2A (\partial_k H_0) \cos(\omega t - \gamma) / \hbar \omega$ is given by
\begin{align}
R_{jk}^{(\gamma)}
&= \frac{2\pi }{\hbar } \frac{ A^2 }{ (\hbar\omega)^2}  \left| \left\langle \mathrm{e} \right| \left( \partial_j H_0 + e^{i \gamma} \partial_k H_0 \right) \left| \mathrm{g} \right\rangle \right|^2 \delta(2d - \hbar \omega) \, .
\end{align}
Similarly, the matrix element evaluates to
\begin{align}
&\left| \left\langle \mathrm{e} \right| \left( \partial_j H_0 + e^{i \gamma} \partial_k H_0 \right) \left| \mathrm{g} \right\rangle \right|^2 
\nonumber \\
&\stackrel{\eqref{eq:sm22}}{=} 4d^2
\bigl( 
\left\langle  \partial_j  \mathrm{g} \middle|\mathrm{e} \right\rangle  \left\langle \mathrm{e} \middle| \partial_j  \mathrm{g} \right\rangle 
+ 
e^{i \gamma} \left\langle  \partial_j  \mathrm{g} \middle|\mathrm{e} \right\rangle  \left\langle \mathrm{e} \middle| \partial_k  \mathrm{g} \right\rangle 
\nonumber \\
&\qquad \qquad + 
e^{-i \gamma}  \left\langle \partial_k  \mathrm{g}  \middle| \mathrm{e}\right\rangle \left\langle \mathrm{e} \middle| \partial_j  \mathrm{g} \right\rangle 
+ 
\left\langle \partial_k  \mathrm{g}  \middle| \mathrm{e}\right\rangle  \left\langle \mathrm{e} \middle| \partial_k  \mathrm{g} \right\rangle  \bigr)
\nonumber \\
&= 4d^2
\bigl( 
g_{jj} + g_{kk}
+ 
e^{i \gamma} \chi_{jk}
+ 
e^{-i \gamma} \chi_{jk}^*
\bigr)
\nonumber \\
&= 4d^2
\bigl( 
g_{jj} + g_{kk}
+ 
2 g_{jk} \cos\gamma 
+ 
F_{jk} \sin\gamma
\bigr).
\end{align}
The different oscillator strengths of the transition rates for linearly polarized absorption are shown in Fig.~\ref{fig:sfig3}.
\subsection*{Three-terminal junction with magnetic flux: Hamiltonian, Weyl nodes, and Chern number}
We briefly discuss the case of $n = 3$ SC terminals with nonzero magnetic flux $\Phi = n \alpha \Phi_0$.
The system is sketched in Fig.~\hyperref[fig:sfig2]{S3(a)}. 
As discussed above, the Hamiltonian takes the form of a pseudo-spin in an effective magnetic field, i.e., $H_0 = \boldsymbol{d} \cdot \boldsymbol{\tau}$ and the Andreev bound states $\varepsilon_\mathrm{e/g} = \pm d$, where $d = |\boldsymbol{d}|$. The expression of the effective magnetic field is given in Eq.~\eqref{eq:magField}.
Gauge invariance allows us to set one SC phase to zero (we set $\varphi_3 = 0$ in the following) and the magnetic flux plays a similar role as one of the SC phases.
The spectrum is shown in Figs.~\hyperref[fig:sfig2]{S3(b){\textendash}S3(f)} for different values of $\alpha$. 
In this system, we find that Weyl nodes $\boldsymbol{\varphi}_\mathrm{W} = (\varphi_{\mathrm{W},1},\varphi_{\mathrm{W},2},\alpha_\mathrm{W})$, for which $d(\boldsymbol{\varphi}_\mathrm{W})=0$, only exist for $-6 \leq m \leq 6$ with $m = \varepsilon_0 / t_0 \Gamma$.
There are four Weyl nodes located at (modulo $2\pi$ in each direction)
\begin{subequations}
	\begin{align}
	\boldsymbol{\varphi}^{(1)}_\mathrm{W} &= \bigl( -2 \pi / 3, \, 2 \pi / 3 , \, \alpha_+(m) \bigr) ,\\
	\boldsymbol{\varphi}^{(2)}_\mathrm{W} &= \bigl( -2 \pi / 3 , \, 2 \pi / 3 , \, \alpha_-(m) \bigr) ,\\
	\boldsymbol{\varphi}^{(3)}_\mathrm{W} &= \bigl( 2 \pi / 3 , \, -2 \pi / 3 , \, - \alpha_-(m) \bigr) ,\\
	\boldsymbol{\varphi}^{(4)}_\mathrm{W} &= \bigl( 2 \pi / 3 , \, -2 \pi / 3 , \, - \alpha_+(m) \bigr) ,
	\end{align}
\end{subequations}
where
\begin{align}
\alpha_{\pm}(m) &= 2 \arctan\left( \frac{\sqrt{3} \pm 2\sqrt{1 - (m/6)^2}}{m/3 -1} \right) .
\end{align}
Each Weyl node $\boldsymbol{\varphi}^{(s)}_\mathrm{W}$, $s = 1,2,3,4$, carries a topological charge $c_s$ which is obtained in the same way as for the four-terminal junction. 
We find $c_{1,4} = \mathrm{sgn}(t_0)$ and $c_{2,3} = - \mathrm{sgn}(t_0)$.
The locations and the charges of the Weyl points are shown in Fig.~\hyperref[fig:sfig2]{S3(g)}.

We define a first Brillouin zone (FBZ) as $(\varphi_1 , \varphi_2, \alpha) \in [0,2\pi)^3$ and calculate the Chern number $C$ as a function of $\alpha$ from the normalized effective field $\boldsymbol{n} = \boldsymbol{d}/d$ via
\begin{align}
C (\alpha) = \frac{1}{4\pi} \int_{0}^{2\pi} \!\!\! \mathrm{d}\varphi_1 \int_{0}^{2\pi} \!\!\! \mathrm{d}\varphi_2 \, \Bigl( \boldsymbol{n} \cdot [ (\partial_{\varphi_1} \boldsymbol{n}) \times (\partial_{\varphi_2} \boldsymbol{n}) ] \Bigr) .
\end{align}
The Chern number is shown in Fig.~\hyperref[fig:sfig2]{S3(h)} and shows topologically nontrivial regions for certain values of $\alpha$. 
Since we integrate over the phases $\varphi_1$ and $\varphi_2$, the finite jumps of $C$ are associated with the values $\alpha_\pm$ and $-\alpha_\pm$. 
In particular for $m = 0$, the four values of a topological phase transition are $\alpha_\mathrm{W} = \pi/6, 5\pi/6, 7\pi/6, 11\pi/6$ in the FBZ.


\begin{thebibliography}{10}
\bibitem{footnote1}
	A similar calculation can be done for the excited state $|\mathrm{e}\rangle$ which contains the same information since the ABS come in particle-hole symmetric pairs.
\bibitem{footnote2}
	A similar figure for the elements of the metric tensor $g$ and its relation to the oscillator strengths for linearly polarized absorptions [Eqs.~\eqref{eq:FGRonephase} and \eqref{eq:offDiagonalOfMetric}] is shown in Fig.~\ref{fig:sfig3}, see Ref. \cite{sMaterial}.
\end{thebibliography}
\end{document}